\documentclass[aps,prb,twocolumn]{revtex4} 

\usepackage{graphicx}
\usepackage{dcolumn}
\usepackage{bm}
\usepackage{amsmath}  

\newcommand{\comment}[1]{}

\begin{document}
\renewcommand{\theequation}{\arabic{section}.\arabic{equation}}

\title{The Paradox of Bose-Einstein Condensation}


\author{Phil Attard}
\affiliation{ {\tt phil.attard1@gmail.com}  July 16, 2023}


\begin{abstract}
The paradox of Bose-Einstein condensation is that phenomena
such as the  $\lambda$-transition heat capacity and superfluid flow
are macroscopic, whereas the occupancy of the ground state is microscopic.
This contradiction is resolved with a simple derivation for ideal bosons
that shows Bose-Einstein condensation is into multiple low-lying states,
not just the ground state.
\end{abstract}


\maketitle

%
%


Bose-Einstein condensation has a contradiction at its heart.
Einstein wrote
in a letter to Paul Ehrenfest (1924),
\begin{center}
\parbox{7.5cm}{
`From a certain temperature on,
the molecules ``condense'' without attractive forces,
that is, they accumulate at zero velocity.' (Balibar 2014)
}\end{center}
Although Einstein was specifically discussing ideal bosons,
in which the energy and the momentum ground states are the same,
Bose-Einstein condensation has ever since been generally considered
as occurring in the energy ground state, even for interacting bosons.

The paradox arises because the size of the ground state
decreases with increasing subsystem size.
Specifically, the spacing of momentum states is $\Delta_p = 2\pi \hbar/L$
(Messiah 1961, Merzbacher  1970),
where $\hbar$ is Planck's constant divided by $2\pi$,
and $L$ is the edge length of the subsystem;
the momentum volume of the ground state is $\Delta_p^3 = (2\pi \hbar)^3/V$,
where $V=L^3$ is the volume of the subsystem.
Since the size of the ground state decreases
with increasing  subsystem size,
the occupancy of the ground state must be an intensive thermodynamic variable:
if the size of the subsystem is doubled,
then both the number of bosons and the number of states
in a given range are also doubled,
leaving the occupancy of each state unchanged.
Mathematically,
ideal bosons have an average  ground state occupancy
$\overline N_{\bf 0}(z)=z/(1+z)$,
where the fugacity $z$ is an intensive variable (see below).
In consequence if Bose-Einstein condensation was indeed into
the ground state then it would not be measurable by any macroscopic method.

However it is widely believed
that Bose-Einstein condensation is a macroscopic phenomena
ever since F. London's (1938) ideal boson analysis
that explained the $\lambda$-transition and superfluid flow
in liquid helium-4 in terms of it.
Since the $\lambda$-transition is signified by the peak
in the heat capacity, which is an extensive thermodynamic variable,
Bose-Einstein condensation must itself be extensive.
Similarly, the fact that  superfluid flow is observable with the naked eye
must mean that Bose-Einstein condensation is also macroscopic in nature.

Hence one has two contradictory interpretations:
On the one hand general thermodynamic arguments
show that the occupancy of the ground state is an intensive variable,
and so by Einstein's definition
that Bose-Einstein condensation is into the ground state,
it must also be intensive and independent of subsystem size.
On the other hand the $\lambda$-transition and superfluid flow
are both macroscopic phenomena,
and in so far as Bose-Einstein condensation is the basis for both
then it must be extensive with the subsystem size.

%
\renewcommand{\theequation}{\arabic{equation}}
%

To resolve this paradox
let us re-analyse the ideal boson treatment of the $\lambda$-transition
of F. London (1938), as set out by Pathria (1972 section~7.1).
For ideal bosons,
the partition function can be written as the product of the sums
over the occupancies of the
single particle momentum states ${\bf a} = \{ a_x,a_y,a_z\} = {\bf n} \Delta_p$,
where ${\bf n}$ is a three-dimensional integer.
Hence the grand potential is given by
(Pathria 1972 section~6.2, Attard 2023a section~7.7)
\begin{eqnarray}
-\beta \Omega & = &
\ln \prod_{\bf a} \sum_{N_{\bf a}=0}^\infty
z^{N_{\bf a}} e^{-\beta N_{\bf a} a^2 /2m }
\nonumber \\ & = &
- \sum_{\bf a} \ln \big[ 1 - z e^{-\beta a^2 /2m } \big].
\end{eqnarray}
Here $\beta = 1/k_\mathrm{B}T$ is the inverse temperature,
$a^2 /2m$ is the kinetic energy
of the single particle momentum state ${\bf a}$,
and $z=e^{\beta \mu}$ is the fugacity,
$\mu$ being the chemical potential.
The average total number of bosons is given by
the usual derivative (Pathria 1972, Attard 2023a)
\begin{equation}
\overline N
= \frac{z \partial(-\beta \Omega )}{\partial z}
= \sum_{\bf a} \frac{ ze^{-\beta a^2/2m }}{1-ze^{-\beta a^2/2m }} .
\end{equation}
The summand is the average momentum state occupancy
$\overline N_{\bf a}$.

Choose a momentum magnitude $a_0$
corresponding to some fraction of the thermal energy, such that
$\nu \equiv \beta a_0^2 /2m < 1$.
The number of momentum states in the neighborhood of the ground state
by this criterion is
$M_0 = 4\pi a_0^3/3\Delta_p^3
= (4\pi/3) (\nu /\pi)^{3/2} V/\Lambda^3 $.
Here $\Lambda \equiv \sqrt{2\pi\hbar\beta/m}$ is the thermal wavelength,
which is of molecular size and which routinely arises from wave function
symmetrization effects (Pathria 1972, Attard 2023a).
The number of states in the neighborhood is  macroscopic
and it increases with increasing subsystem size.

For $\nu$ chosen small enough
we may replace $e^{-\beta a^2/2m } \Rightarrow 1$ for $a \le a_0$.
With this the sum over states for the average number of bosons
may be split into two,
the first containing constant terms,
and the second approximated by a continuum integral,
\begin{eqnarray}
\overline N
& \approx &
\sum_{\bf a} \,\!^{( a \le a_0)}  \frac{z }{1-z}
+ \sum_{\bf a} \,\!^{( a > a_0)}
\frac{ ze^{-\beta a^2/2m }}{1-ze^{-\beta a^2/2m }}
\nonumber \\ & \approx &
M_0  \frac{z }{1-z}
+
\frac{1}{\Delta_p^3}
\int_{a_0}^\infty \mathrm{d} a\; 4\pi a^2\,
\frac{ ze^{-\beta a^2/2m }}{1-ze^{-\beta a^2/2m }}
\nonumber \\ & \approx &
M_0 \frac{z }{1-z}
+
\frac{1}{\Delta_p^3}
\int_{0}^\infty \mathrm{d} a\; 4\pi a^2\,
\frac{ ze^{-\beta a^2/2m }}{1-ze^{-\beta a^2/2m }}
\nonumber \\ & = &
M_0  \frac{z }{1-z}
+ V \Lambda^{-3} g_{3/2}(z)
\nonumber \\ & \le &
M_0  \frac{z }{1-z}
+ V \Lambda^{-3} \zeta(3/2) , \;\; T \alt T_\lambda.
\end{eqnarray}
The second term, which is the number of uncondensed bosons,
involves the Bose-Einstein integral,
$g_{n}(z)
= \Gamma(n)^{-1} \int_0^\infty \mathrm{d}x\;
x^{n-1} z e^{-x}/[1-z e^{-x}]
= \sum_{l=1}^\infty z^l l^{-n}$
(Pathria 1972 section 7.1, Attard 2023a section~8.2.2).
The final equality holds in the vicinity of the $\lambda$-transition,
with the maximum density of uncondensed ideal bosons being
$\rho^\mathrm{id}_* \Lambda^3 \le g_{3/2}(1) = \zeta(3/2) = 2.612\ldots$.
When the actual density exceeds this value,
the additional bosons are given by the first term,
and Bose-Einstein condensation is said to occur.

In the third equality  the integral has been extended to the origin,
with the maximum error at $z=1$ being
$\Delta_p^{-3} \times a_0 \times 4\pi a_0^2 /(\beta a_0^2/2m)
= 4 (\nu/\pi)^{1/2} V/\Lambda^3$.
This increases the number of uncondensed bosons by a factor of $1+\surd\nu$,
which error can be neglected.

The conventional derivation (F. London 1938, Pathria 1972 section~7.1)
sets $M_0=1$, which limits the condensed bosons solely to the ground state.
In this case the number of condensed bosons equals
the number of ground state bosons,
$ \overline N_{000}=z/(1-z)$, which is intensive.
In the present analysis
the $M_0 = (4\pi/3) (\nu /\pi)^{3/2} V/\Lambda^3 $
states in the neighborhood of the ground state
are occupied by condensed bosons.
This number of states grows with the size of the subsystem
while the occupancy of each state remains unchanged.
Even for an error of say 1\%, $\nu \sim 10^{-4}$,
since $V/\Lambda^3 $ is on the order of Avogadro's number
the number of condensed states  is macroscopic.

The original criterion for the $\lambda$-transition
given by F. London (1938)
also holds for the present analysis:
condensation occurs when the saturated liquid density and thermal wave length
exceed the number of uncondensed bosons given by the continuum integral,
$\rho \Lambda^3 > \zeta(3/2)$.
For $^4$He at the measured liquid saturation density
this corresponds to $T_\lambda^\mathrm{id} = 3.13$\,K,
which is close to the measured value, $T_\lambda = 2.19$\,K.

Obviously the virtue of ideal boson analysis is qualitative
rather than quantitative.
It reveals the physical basis of the phenomenon,
and the approximate agreement with reality must be regarded as a bonus.

The present result has the interpretation
that states within about the thermal energy
of the  ground state contain condensed bosons (ie.\ are highly occupied),
and uncondensed bosons inhabit states beyond the thermal energy
(ie.\ such states are empty or sparsely occupied).
This makes more physical sense than Einstein's (1924) and F. London's (1938)
assertion that bosons condense solely into the ground state.
The present analysis fills the lacuna in Pathria's (1972 section~7.2)
derivation of the ideal boson result where his justification
for adding the ground state contribution to the continuum integral
is a little lame, and it extends that derivation beyond the ground state.
The present result resolves
the problem of the missing latent heat at the $\lambda$-transition
(if a macroscopic number of bosons condensed into the ground state
at the transition
then there would be a discontinuous change in energy).
It also makes sense for superfluid flow,
which necessarily involves bosons with non-zero momentum.
This result is consistent
with the discussion in Attard (2023a chapters 8 and 9)
for the $\lambda$-transition,
although the simulated transition temperature
for $^4$He bosons interacting with the Lennard-Jones pair potential,
is based on ground state condensation only (Attard 2023a section~8.5).
The result is also consistent with the recent calculation and explanation
for the superfluid viscosity (Attard 2023b).
The present analysis complements these earlier arguments with mathematical rigor,
and yields a consistent picture of the $\lambda$-transition,
superfluidity, and Bose-Einstein condensation.

\section*{References}


\begin{list}{}{\itemindent=-0.5cm \parsep=.5mm \itemsep=.5mm}


\item 
Attard  P 2023a
\emph{Entropy Beyond the Second Law.
Thermodynamics and Statistical Mechanics
for Equilibrium, Non-Equilibrium, Classical, And Quantum Systems}
(Bristol: IOP Publishing, 2nd edition)

\item 
Attard  P 2023b
Quantum Stochastic Molecular Dynamics Simulations
of the Viscosity of Superfluid Helium
arXiv:2306.07538

\item 
Balibar S (2014)
Superfluidity: How Quantum Mechanics Became Visible.
In:
Gavroglu, K. (eds)
\emph{History of Artificial Cold, Scientific,
Technological and Cultural Issues.}
Boston Studies in the Philosophy
and History of Science, {\bf 299}
(Dordrecht: Springer)

\item  
London F 1938
The $\lambda$-Phenomenon of Liquid Helium and the Bose-Einstein Degeneracy
\emph{Nature} {\bf 141} 643

\item 
Merzbacher E 1970
\emph{Quantum Mechanics} 2nd edn
(New York: Wiley)

\item 
Messiah A 1961 \emph{Quantum Mechanics}
(Vol 1 and 2) (Amsterdam: North-Holland)

\item
Pathria R K 1972
\emph{Statistical Mechanics} (Oxford: Pergamon Press)

\end{list}


\end{document}